\documentclass{PoS}
\usepackage{url}
\usepackage[dvipsnames]{xcolor}
\usepackage{graphicx}

\title{Novel Approaches to High-Power Proton Beams}

\ShortTitle{Novel Approaches to High-Power Proton Beams}

\author{\speaker{J. Eldred} \thanks{Operated by Fermi Research Alliance, LLC under Contract No. DE-AC02-07CH11359 with the United States Department of Energy.} \\
        Fermi National Accelerator Laboratory, Batavia, Illinois 60510, USA \\
        E-mail: \email{jseldred@fnal.gov}}

\abstract{The potentially realizable beam power at the Fermilab long-baseline neutrino program has motivated a reinvigorated design and optimization effort for a rapid-cycling synchrotron (RCS) intensity upgrade of the Fermilab proton complex. We examine areas of technological development with the potential for high-impact on the Fermilab RCS design - low-loss slip-stacking, advanced neutrino-target R\&D, laser-stripping H$^{-}$ injection, fast-ramping super-ferric magnets, nonlinear integrable optics, electron lens devices, and next-generation halo monitors. A brief overview is given on the Fermilab Accelerator Science \& Technology (FAST) facility, where the latter three technologies will undergo comprehensive beam tests.}

\FullConference{The 21st International Workshop on Neutrinos (NuFact2019)\\
        25-31 August 2019\\
        Daegu, South Korea}

\begin{document}

\section{Introduction}

With the advent of the Deep Underground Neutrino Experiment (DUNE) and Long Baseline Neutrino Facility (LBNF) program~\cite{Yu,DUNEidr} there is strong motivation for a 2.4~MW beam power upgrade of the Fermilab proton facility. The Proton Improvement Plan II (PIP-II)~\cite{PIP2} is a set of upgrades and improvements to the Fermilab accelerator complex to achieve 1.2 MW beam power at 120 GeV. A recent design study~\cite{Eldred2019} proposes to enable 2.4~MW Main Injector (MI) beam power by replacing the Fermilab Booster with a new rapid-cycling synchrotron (RCS). The proposal is unique in that it envisions multi-MW slip-stacking, although alternative approaches are discussed. Table~\ref{Upgrades} shows the present, planned, and proposed operational parameters for the Fermilab Proton Complex.

\begin{table}[htp]
\centering
\begin{tabular}{| l | l | l | l | l | l |}
\hline
~ & Present & PIP-II & RCS \\
\hline
Linac Current & 24~mA & 2~mA & 2~mA \\
Linac Energy & 0.4~Gev & 0.8~GeV & 1.0~GeV \\
\hline
Booster/RCS Energy & 8~GeV & 8~GeV & 11~GeV \\
Booster/RCS Circumference & 474~m &474~m & 600~m \\
Booster/RCS Emittance & 16~mm~mrad & 16~mm~mrad & 24~mm~mrad \\
Booster/RCS Ramp Rate & 15~Hz & 20~Hz & 15~Hz \\
Booster/RCS Intensity & 4.5~$\times~10^{12}$ & 6.5~$\times~10^{12}$ & 25~$\times~10^{12}$ \\
Booster/RCS Available Power & 25~kW & 80~kW & 440~kW \\
\hline
Main Injector Cycle Time & 1.33~s & 1.2~s & 1.8~s \\
Main Injector Intensity & 54 e12 & 78 e12 & 225 e12 \\
Main Injector Power & 0.75~MW & 1.2~MW & 2.4~MW \\
\hline
Year & 2019 & $\sim$2027 & $\sim$2032 \\
\hline
\end{tabular}
\caption{Beam parameters for current Fermilab Proton Complex, planned PIP-II upgrade, and RCS upgrade proposed in \cite{Eldred2019}. Timeline for Main Injector power upgrades following Deep Underground Neutrino Experiment (DUNE) Interim Design Report~\cite{DUNEidr}.}
\label{Upgrades}
\end{table}

This 11~GeV RCS design study proposed to achieve the 2.4~MW benchmark ``without deviating from well-tested accelerator design principles''. However advanced accelerator technology can enhance the performance of this RCS design, and that of high-power hadron rings generally. In this paper we will use this RCS design provides as a case study of the underlying physics challenges that next-generation technology might address.

A new 11~GeV permanent-magnet storage ring (SR) to replace the Fermilab Recycler is recommended to be constructed in tandem with the proposed 11~GeV Fermilab RCS.
The facility would reach 2~MW during beam commissioning when five batches of 25$~\times~10^{12}$ protons could be conventionally accumulated in the SR and sent to the Main Injector. Then 2.4~MW beam power or higher could then be achieved either by commissioning slip-stacking in the SR or by achieving more aggressive RCS intensity targets which by be enabled by emerging technology.

Table~\ref{SSHI} shows the required intensity under these various RCS operation scenarios with use of a storage ring accumulator. The scenarios can be compared on the basis of the machine performance requirements but also entail a significant variation in the available 11~GeV beam power.

\begin{table}[htp]
\centering
\resizebox{\linewidth}{!}{%
\begin{tabular}{| l | l | l | l | l |}
\hline
\begin{tabular}{l} RCS Scenarios \\ ~ with Storage Ring \end{tabular} & \begin{tabular}{l} {\bf 2.4 MW } with \\ ~ Slip-stacking \end{tabular} &
\begin{tabular}{l} {\bf 3.6 MW } with \\ ~ Slip-stacking \end{tabular} &
\begin{tabular}{l} {\bf 2.4 MW } with \\ ~ Conventional \end{tabular} &
\begin{tabular}{l} {\bf 3.6 MW } with \\ ~ Conventional \end{tabular} \\
\hline
RCS Intensity & 17 e12 & 25 e12 & 30 e12 & 45 e12 \\ \hline
Number of Batches & 9 & 9 & 5 & 5 \\ \hline
MI Intensity & 150 e12 & {\color{red} 225 e12} & 150 e12 & {\color{red} 225 e12} \\ \hline
RCS Tune-shift & {\color{ForestGreen} 0.16-0.24} & 0.24-0.35 & {\color{red} 0.29-0.42} & {\color{red} 0.43-0.63} \\ \hline
Available RCS Power & 225~kW & 330~kW & {\color{ForestGreen} 570~kW} & {\color{ForestGreen} 860~kW} \\ \hline
\end{tabular}}
\caption{Scenarios for 2.4~MW \& 3.6~MW proton complex with new RCS and storage ring, using slip-stacking accumulation or conventional boxcar stacking. The table also indicates the 11~GeV RCS beam power that is available concurrently with 120~GeV Main Injector operation.}
\label{SSHI}
\end{table}

\section{High-Impact Technologies}

The RCS scenarios outlined in Table~\ref{Upgrades} and Table~\ref{SSHI} entail a factor 3-4 increase in Main Injector pulse intensity compared to present operation of long-baseline neutrino program. Davenne et al.~\cite{Davenne} provides a conceptual 2.4~MW target design based on segmented beryllium and optimized for production of neutrino beams from a 160~$\times~10^{12}$ proton pulse.

Designing and operating the multi-MW generation of neutrino targets will critically rely on radiation damage R\&D for low-Z high-temperature materials such as carbon and beryllium. The Radiation Damage In Accelerator Target Environments (RaDIATE) collaboration~\cite{Senor} was developed to study a broad suite of accelerator target, dump, window, and collimator materials under extensive radiation damage and thermal shock. As part of the RaDIATE collaboration, accelerator relevant materials are exposed to high radiation dose at the Brookhaven Linac Isotope Producer (BLIP) facility where they can be subject to examination and testing procedures. CERN's High-Radiation to Materials (HiRadMat) facility enables the post-irradiation materials to then subsequently undergo thermal shock testing from a high-energy accelerator beam.

With present operational technology, protons would be accumulated in the Fermilab RCS via charge-stripping foil injection. The linear optics of the RCS~ \cite{Eldred2019} are heavily constrained by the foil-heating limits (large-betas) and particle loss due to scattering off the injection foil (phase-advance into collimators). However, the combined us of use of high-power lasers and magnetic fields has an emerged as an alternative approach to stripping H$^{-}$ beams. In \cite{Cousineau}, 1~GeV~H$^{-}$ are stripped with 95\% stripping efficiency over a 10~$\mu$s timescale with an ultraviolet laser delivering 1~MW peak power. The 10 $\mu$s timescale is not suitable for a multi-millisecond injection but the paper articulates that ``a doubly resonant optical cavity scheme is being developed to realize a cavity enhancement of burst-mode laser pulses'' (see \cite{Rakhman}).

In \cite{Piekarz}, a fast-ramping super-ferric dipole is demonstrated based on high-temperature superconductor (HTS) REBCO wires. The prototype magnet applies the dipole field to two vacuum chambers to make full use of the return flux of the wire. This technology could be applied to a compact RCS design to use RF power more efficiently. For example, a 300~m RCS could extract 2$\times$15 e12 protons every 5~Hz to a storage ring as part of a 2.4~MW Main Injector facility with only 1/6 of the required RF power in the RCS. Alternatively, the technology could be applied to extend the energy reach of an RCS design. For example, a 500~m RCS could extract 2$\times$25 e12 protons at 22~GeV every 7.5~Hz in order to inject above the Main Injector transition energy and enable 2.4~MW beam power at 120~GeV. High-field, fast-ramping super-ferric magnets open up a larger machine-design parameter space then previously considered in \cite{Eldred2019}.

\section{FAST/IOTA Research}

The Integrable Optics Test Accelerator (IOTA) and Fermilab Accelerator Science \& Technology (FAST) facility were created for advanced accelerator R\&D and beam physics research~\cite{AntipovIOTA}. The IOTA ring was commissioned for electron beam studies earlier this year and will be commissioned for proton beam studies starting in 2020~\cite{ValishevIOTA}. Nonlinear integrable optics and electron lens space-charge compensation are two technologies under development at FAST/IOTA to be tested for their application to intense hadron rings.

Nonlinear integrable optics is an innovation in acceleration design to provide immense nonlinear focusing without generating parametric resonances~\cite{Danilov}. Strong nonlinear focusing could significantly enhance the performance of an RCS by mitigating halo formation and damping collective instabilities~\cite{Eldred2019}.

The canonical formulation of integrable lattice design expresses the single-particles dynamics with precisely defined phase advances between specialized nonlinear sections~\cite{Danilov}. The perturbation of the system under space-charge~\cite{Eldred18} \& chromatic effects~\cite{Webb} has been investigated. With several superperiodic nonlinear cells, the integrable lattice design seems to be compatible with extreme nonlinearity under realistic tune-shifts; however a systematic study has not been conducted for tune-shift tolerances as a function of nonlinearity. Separately, a framework for nonlinear integrable lattices based on distributed thin nonlinear elements has also been proposed~\cite{Baturin}.

The electron lens is a versatile particle accelerator device with applications in beam-beam compensation~\cite{ShiltsevTev}, collimation~\cite{Mirachi}, nonlinear focusing~\cite{ShiltsevLHC}, and space-charge compensation. Recent work~\cite{Stern} examines the performance of electron lens space-charge compensation in the extreme space-charge case, but further work remains to extend the work to a constrained lattice environment.

\section{Beam Profile \& Halo Diagnostics}

Intense proton rings often require precise management of halo particles, even while those halo particles are very sensitive to the transverse distribution of the charge-dominated beam~\cite{Jeon}. Accordingly, successful commissioning and operation of intense proton rings then requires proton diagnostics which provide information about the evolution of the beam profile over the cycle as well as the population of lost or outlying particles. Ionization profile monitors are the traditional ring diagnostic to measure proton beam profiles, such as those that have been implemented at the Fermilab Booster~\cite{Amundson}

A related diagnostic device, the gas-sheet profile monitor, has several advantages over ionization-profile monitors. The high-velocity gas flow is collected to minimize the gas load on the vacuum system. The pressure of the gas-sheet can be varied in order to change continuously from a minimally destructive diagnostic to a turn-by-turn profile diagnostic with mm-scale resolution. A gas-sheet profile monitor is being developed as a diagnostic for the Fermilab IOTA ring~\cite{Szustkowski}, where it will observe a high-space proton beam in highly nonlinear potentials. In the future, specialized gas monitors might serve as a halo diagnostic by intercepting the beam tails with a narrow high-pressure gas jet.

At the Fermilab Booster, photo-multiplier tube (PMT) scintillator detectors have been placed at several locations outside the Booster ring to serve as fast loss-monitors. The fast loss monitors are synchronized to accelerator control signals to within a few nanoseconds, allowing particle loss mechanisms to be diagnosed from their temporal signature~\cite{Bhat}. The fast loss monitors sample an unknown fraction of the lost particles, and therefore can only be used to infer relative changes in loss patterns. In the future, scintillator-based particle detectors might be directly integrated into the design of beam collimators to more systematically measure lost halo particles. 

\section{Overview}

In this paper we've introduces some variations in Fermilab RCS scenarios that constitute important trade-offs between space-charge limits, intensity requirements, and delivered beam power. High-power neutrino targetry directly impacts the realizable Main Injector pulse intensity and beam power, while laser-stripping technology directly impacts the RCS intensity and overall RCS design. Fast-ramping HTS super-ferric magnets have an unexplored potential to broaden the machine parameters for RCS design.

The Fermilab AST/IOTA Facility is developing advanced particle accelerator technology, which may find a direct application in the proposed Fermilab RCS. Nonlinear integrable optics can reduce the number of particles requiring collimation and raise the threshold for collective instabilities (without introducing new nonlinear resonances). Electron lens devices have many valuable applications for high-intensity particle accelerators, but a new application for direct space-charge compensation will be investigated in simulation and experiment over the next several years. And finally, the next generation of proton beam diagnostics will be capable of delivering higher dynamic range with faster time resolution.


\begin{thebibliography}{99}

\bibitem{Yu} J.~Yu, \emph{Status and Plans of DUNE}, in \emph{Proceedings of NUFACT2019},  Daegu, South Korea, August 2019.

\bibitem{DUNEidr} B.~Abi et al. (DUNE Collaboration), \emph{The DUNE Far Detector Interim Design Report Volume 1: Physics, Technology and Strategies}, Fermilab Technical Note, Fermilab-Design-2018-02, 2018 arXiv:1807.10334 [{\tt physics.ins-det}].

\bibitem{PIP2} V.~Lebedev et al. (PIP-II Collaboration), {\sl The PIP-II Conceptual Design Report}, FERMILAB-TM-2649-AD-APC (2017).

\bibitem{Eldred2019} J. Eldred, V.~Lebedev and A.~Valishev, \emph{Rapid-cycling synchrotron for multi-megawatt proton facility at Fermilab}, \emph{JINST} {\bf 14} (2019) P07021.

\bibitem{Davenne} T.~Davenne et al., \emph{Segmented beryllium target for a 2 MW super beam facility}, \emph{Phys. Rev. Accel. Beams} {\bf 21} (2018) 053001.

\bibitem{Senor} D.~Senor, \emph{Radiation Damage Experiments Update from the RaDIATE Collaboration}, in \emph{Proceedings of NUFACT2019}, Daegu, South Korea, August 2019.

\bibitem{Cousineau} S.~Cousineau et al., \emph{High efficiency laser-assisted H- charge exchange for microsecond duration beams}, \emph{Phys. Rev. Accel. Beams} {\bf 20} (2017) 120402. arXiv:1805.04481 [{\tt physics.acc-ph}]

\bibitem{Rakhman} A.~Rakhman et al., \emph{Multifunctional optical correlator for picosecond ultraviolet laser pulse measurement}, \emph{Applied Optics} {\bf 53} 7603 (2014). 

\bibitem{Piekarz} H.~Piekarz et al., \emph{Record fast-cycling accelerator magnet based on HTS conductor}, \emph{NIM A} {\bf 943} 162490 (2019).

\bibitem{AntipovIOTA} S. Antipov et al., \emph{IOTA (Integrable Optics Test Accelerator): Facility and Experimental Beam Physics Program}, \emph{JINST} {\bf 12} T03002 (2017). arXiv:1612.06289 [{\tt physics.acc-ph}]

\bibitem{ValishevIOTA} A.~Valishev, \emph{IOTA/FAST Status, First Run and Outlook}. in \emph{FAST/IOTA Collaboration Meeting}.

\bibitem{Danilov} V.~Danilov and S.~Nagaitsev, \emph{Nonlinear Accelerator Lattices with One and Two Analytic Invariants}, \emph{ Phys. Rev. ST Accel. Beams} {\bf 13} (2010) 084002. arXiv:1003.0644 [{\tt physics.acc-ph}]

\bibitem{Eldred18} J. Eldred and A. Valishev, \emph{Simulation of Integrable Synchrotron with Space-Charge and Chromatic Tune-Shifts}, in \emph{Proceedings of IPAC'18}, TUPAF073, Vancouver, British Columbia, Canada, April 2018. arXiv:1805.02134 [{\tt physics.acc-ph}]

\bibitem{Webb} S.~D.~Webb, N.~Cook and J.~Eldred, \emph{Effects of Synchrotron Motion on Nonlinear Integrable Optics}, in \emph{Proceedings of IPAC'18}, THPAF067, Vancouver, British Columbia, Canada, April 2018.

\bibitem{Baturin} S.~S.~Baturin, \emph{Hamiltonian preserving nonlinear optics}. arXiv:1908.03520 [{\tt physics.acc-ph}].

\bibitem{ShiltsevTev} V.~Shiltsev et al., \emph{Experimental Demonstration of Colliding-Beam-Lifetime Improvement by Electron Lenses}, \emph{Phys. Rev. Lett.}, {\bf 99}, (2007) 244801. 

\bibitem{Mirachi} D.~Mirachi et al., \emph{Hollow electron-lens assisted collimation and plans for the LHC}, in \emph{Proceedings of HB2018}, Daejeon, Korea, June 17-22, 2018.

\bibitem{ShiltsevLHC} V.~Shiltsev et al., \emph{Landau Damping of Beam Instabilities by Electron Lenses}, \emph{Phys. Rev. Lett.}, {\bf 119}, (2017) 134802. arXiv:1706.08477 [{\tt physics.acc-ph}].

\bibitem{Stern} E.~Stern, \emph{Simulation of Space Charge Compensation with Electron Lenses}, in \emph{Proceedings of Space Charge 2019}, CERN, Geneva, Switzerland, November 2019.

\bibitem{Jeon} D.~Jeon, \emph{Space-charge particle resonances and mode parametric resonances}, in \emph{Proceedings of NUFACT2019}, Daegu, South Korea, August 2019.
    
\bibitem{Amundson} J.~Amundson, \emph{Calibration of the Fermilab Booster ionization profile monitor}, \emph{Phys. Rev. Accel. Beams} {\bf 6} (2003) 102801.

\bibitem{Szustkowski} S.~Szustkowski, \emph{Development of a Gas Sheet Beam Profile Monitor for IOTA} in \emph{Proceedings of IPAC'18}, WEPAL065, Vancouver, British Columbia, Canada, April 2018.

\bibitem{Bhat} C.~M.~Bhat et. al., \emph{Foil Scattering Model for Fermilab Booster}, in \emph{Proceedings of NAPAC'19}, WEYBB3, Lansing, Michigan, U.S.A., October 2019.

\end{thebibliography}
\end{document}